\documentclass[referee, traditabstract]{aa} %
\usepackage{graphicx}
\usepackage{natbib}
\bibpunct{(}{)}{;}{a}{}{,}
\usepackage{amsmath}
\usepackage{layouts}
\usepackage{color}
    \definecolor{Blue}{rgb}{0.0,0.0,1.0}
    \definecolor{Red}{rgb}{1.0,0.0,0.0}
    \definecolor{Green}{rgb}{0.0,1.0,0.0}

\begin{document}
\title{Mass estimate of the Swift J 164449.3+573451\\ supermassive black hole \\based on
the 3:2 QPO resonance hypothesis}
%
\author{       M.A. Abramowicz\inst{1, 2}
\and           F.K. Liu\inst{3}
}
\institute{    Copernicus Astronomical Center, ul. Bartycka 18, PL-00-716
               Warszawa, Poland
\and          Physics Department, Gothenburg University,
               SE-412-96 G{\"o}teborg, Sweden
                 \\ \email{marek.abramowicz@physics.gu.se}
\and           Astronomy Department, Peking University, Beijing 100871, China
                  \\ \email{fkliu@pku.edu.cn}}
   \date{Received ????; accepted ????}
\abstract {A dormant Swift source J 164449.3+573451 (Sw~164449+57)recently experienced a
powerful outburst, caused most probably by a tidal disruption of a star by the supermassive
black hole at the center of the source. During the outburst, a quasi periodic oscillation
(QPO) was detected in the observed X-ray flux from Sw~164449+57. We show that if the
observed QPO belongs to a ``3:2 twin peak QPO'' (with the second frequency not observed),
the mass of the black hole in Sw~164449+57 is rather low, $M \sim 10^5 M_\odot$, and the
source belongs to a class of intermediate mass black holes. The low mass of the source has
been pointed out previously by several authors.}
\authorrunning{Abramowicz and Liu}\titlerunning{Mass of Sw~164449+57}
\maketitle
The Swift source Sw~164449+57 at redshift z=0.3543 is believed to be a super-massive dormant
black hole at the center of an inactive galaxy. During a recent X-ray outburst (probably
activated by a tidal disruption of a star), the source was in many respects similar to a
small-scale blazar; see e.g. Bloom et al. (2011). In particular, it displayed a relativistic
jet \citep{Burrows-2011,Zauderer-2011}. {If one accepts (as we do) theoretical and
observational arguments that link relativistic jets with a high black hole spin
\citep[discussed by][and many other authors]{Narayan-2012}, then the presence of a jet Sw
164449+57 suggests that the black hole in Sw 164449+57 is rotating rather rapidly ($a
> 0.6$, say). However, one should also note that this evidence of jets being
powered by the black hole rotation is not unanimously accepted. Using much of the same data,
\cite{Fender-2010} came to very different conclusions --- that even the relatively simple
radio-loud versus radio-quiet dichotomy in AGN may be due to the AGN states rather than to
spins. In this context it is also relevant to note that \cite{McKinney-2012} discuss the
possibility of exciting QPOs through a disk-jet coupling.}

During the outburst, \cite{Reis-2012} detected a firm (statistically significant)
QPO\footnote{ It should be mentioned that \cite{Miller-2011a} note the possibility of a QPO
in Swift J1644.} with the centroid frequency $f_{\rm obs} = 4.8(1+z)$ mHz. In this short
article we examine the possibility that this frequency may correspond to a lower (or upper)
frequency of the ``twin peak'' QPOs in which the two frequencies are in the 3:2 ratio. Such
twin peaks are observed in several microquasars and other black hole sources (see e.g.
\cite{Torok-2005}).

It has been argued by \cite{Kluzniak-2001} that the phenomenon of the 3:2 twin peak QPOs in
the black hole sources is due to a nonlinear parametric resonance in two eigenmodes of
accretion disk oscillations. According to the simplest version of the 3:2 resonance model,
the observed QPO twin peak frequencies should be identified with the vertical epicyclic and
radial epicyclic frequencies, which in the Kerr geometry are given by
\begin{eqnarray} \label{epicyclic}
f_{\rm ver} &=& \frac{\Omega_{\rm K}}{2\pi}
\left[ 1 - 4a x^{-3/2} + 3 a^2 x^{-2} \right]^{1/2},
\label{vertical} \\
%
f_{\rm rad} &=& \frac{\Omega_{\rm K}}{2\pi}
\left[ 1 - 6x^{-1} + 8a x^{-3/2} - 3 a ^2 x^{-2} \right]^{1/2},
\label{radial} \\
%
\Omega_{\rm K} &=& \left( \frac{GM}{r^3}\right)^{1/2}
\left[ 1 +x^{-3/2} a \right]^{-1},
\label{keplerian}
\end{eqnarray}
where $M$ is the black hole mass, $a$ its dimensionless spin ($0 \le \vert a \vert \le 1$),
and the dimensionless radial coordinate is defined by
\begin{equation} \label{radius-dimensionless}
x = \frac{r}{GM/c^2}.
\end{equation}
Here $G$ is the Newtonian gravitational constant and $c$ the speed of light. The 3:2
epicyclic resonance occurs at the ``resonance radius'' $x_{3:2} = x_{3:2}(a)$, defined by
the condition
\begin{equation} \label{radius-resonance}
\frac{3}{2} = \left[ \frac{1 - 4a (x_{3:2})^{-3/2} + 3 a^2 (x_{3:2})^{-2}}{1 -
6(x_{3:2})^{-1} + 8a (x_{3:2})^{-3/2} - 3 a ^2 (x_{3:2})^{-2}} \right]^{1/2}.
\end{equation}
For a nonrotating black hole ($a=0$), this implies $x_{3:2}(0) = 54/5$, which is
approximately twice the radius of ISCO.

\begin{figure}[h!]
\begin{center}
\includegraphics[width=0.499\textwidth]{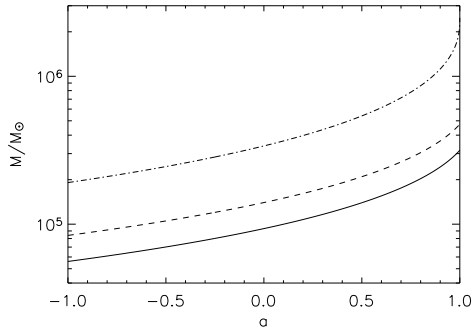}
\end{center}
\caption{Mass estimate for the supermassive black hole in the Swift
  source Sw~164449+57, based on the assumption that the observed QPO frequency
  $f_{\rm obs} = 4.8(1+z)$ mHz corresponds either to the radial epicyclic
  (solid line), $f_{\rm obs} = f_{\rm rad}$, or to the vertical
  epicyclic (dashed line), $f_{\rm obs} = f_{\rm ver}$, frequency,
  in accordance with the twin peak QPO 3:2 resonance model. For
  a comparison, we also show the mass estimate based on an assumption
  that $f_{\rm obs} = \Omega_K/2\pi$ is the Keplerian frequency at
  ISCO (dash-dotted line).
  } \label{fig-mass-estimate}
\end{figure}

The epicyclic resonance hypothesis allows for an accurate estimate of the black hole mass
and spin, as first discussed by \cite{Abramowicz-2001} for the microquasar GRO J1655-40.
Applying this idea to Sw 164449+57, we first assume that $f_{\rm obs} = 4.8(1+z)$ mHz
corresponds to the {\it lower} of the twin peak frequencies, i.e. to the {\it radial}
epicyclic frequency in the Klu{\'z}niak \& Abramowicz resonance
model\footnote{\label{footnote-f2} \cite{Reis-2012} also report a less certain QPO with the
centroid frequency about $f_2 \sim 6.2(1+z)$ mHz. They give no estimate of error in the
determination of this value, but if it is about 16\%, then it could be that $f_2/f_{\rm obs}
\sim 3/2$, which would strengthen the case for $f_{\rm obs} = f_{\rm rad}$.}. After a few
lines of simple algebra we may write, in this case,
\begin{eqnarray} \label{hypothesis-radial}
\frac{M}{M_\odot} &=& A \frac{[1 - 6(x_{3:2})^{-1} + 8a (x_{3:2})^{-3/2} - 3 a ^2
(x_{3:2})^{-2}]^{1/2}}{
(x_{3:2})^{3/2} + a } \nonumber \\
A &=& \frac{c^3}{2\pi G f_{\rm obs} M_{\odot}} = 4.97
  \times 10^6 ,
\end{eqnarray}
where $M_\odot$ is the solar mass. Similarly, if we assume the other possibility, i.e. that
$f_{\rm obs} = 4.8(1+z)$ mHz corresponds to the {\it upper} of the twin peak frequencies, i.e. to
the {\it vertical} epicyclic frequency, we may write
\begin{equation} \label{hypothesis-vertical}
\frac{M}{M_\odot} = A \frac{[1 - 4a (x_{3:2})^{-3/2} + 3 a^2 (x_{3:2})^{-2}]^{1/2}}{
(x_{3:2})^{3/2} + a }.
\end{equation}
Because for a non rotating black hole it is $x_{3:2}(0) = 54/5$, in this case from Eqs.
(\ref{hypothesis-radial}) and (\ref{hypothesis-vertical}) it follows that
\begin{equation} \label{mass-non-rotating}
\frac{M}{M_\odot} =
\begin{cases}
9.34 \times 10^4 &\mbox{if } f_{\rm obs} = f_{\rm rad}, \\
1.40\times 10^5 & \mbox{if } f_{\rm obs} = f_{\rm ver},
\end{cases}
~~ {\rm for}~~ a = 0.
\end{equation}
Similarly, one may calculate that in the two extreme cases, corresponding to $a = \pm 1$. We
consider both the corotating ($a \ge 0$) and counter-rotating ($a \le 0$) cases, because the
star may come to the vicinity of Sw 164449+57 from a random direction before being tidally
disrupted, and the stellar debris can form either a corotating or a counter-rotating
accretion disk. We have
\begin{eqnarray}
\label{mass-max-rotating-plus}
\frac{M}{M_\odot} &=&
\begin{cases}
3.16 \times 10^5  &\mbox{if } f_{\rm obs} = f_{\rm rad}, \\
4.74 \times 10^5 & \mbox{if } f_{\rm obs} = f_{\rm ver},
\end{cases}
~~ {\rm for}~~ a = +1,
%
\\
%
\frac{M}{M_\odot} &=&
\begin{cases}
5.60 \times 10^4 &\mbox{if } f_{\rm obs} = f_{\rm rad}, \\
8.40 \times 10^4 & \mbox{if } f_{\rm obs} = f_{\rm ver},
\end{cases}
~~ {\rm for}~~ a = -1.
\end{eqnarray}
Mass estimates for the case of a rotating black hole for any value of the spin in Sw
164449+57 are given in Figure~\ref{fig-mass-estimate}. Figure~\ref{fig-mass-estimate}
suggests that if the black hole in Sw 164449+57 is rapidly co-rotating with $a > 0.6$, the
mass should be in the range $1.6\times 10^5 < M/M_\odot < 3.2\times 10^5$ if $f_{obs} =
f_{rad}$ or $2.3\times 10^5 < M/M_\odot < 4.7\times 10^5$ if $f_{obs} = f_{ver}$. While, if
the black hole is counter-rotating with $a < -0.6$, the mass would be in the range $5.6
\times 10^4 < M/M_\odot < 6.7 \times 10^4$ if $f_{obs} = f_{rad}$ or $8.4 \times 10^4 <
M/M_\odot < 1.0\times 10^5$ if $f_{obs} = f_{ver}$.

No direct measurement of the mass of the black hole in Sw 164449+57 has been reported. If
the outburst was powered by a tidal disruption of a solar-type main sequence star, the black
hole mass must be $M \la 10^8 M_\odot$ (\cite{Rees-1988}). If a white dwarf was disrupted,
as suggested by \cite{Krolik-2011}, the black hole mass would correspond to the intermediate
mass range,  $M\la 10^5 M_\odot$ {(but see also \cite{Krolik-2012})}. Because the host
galaxy of Sw 164449+57 is not resolved and the host morphology is unknown, the bulge
luminosity cannot be determined. Therefore, the empirical relation (black hole mass -- bulge
luminosity) (\cite{Magorrian-1998}) cannot be used directly to estimate the black hole mass.
By employing the relation of black hole mass and bulge luminosity to the total luminosity of
host galaxy, an upper limit of black hole mass $M \la 10^7 M_\odot$ has been given in the
literature; see \cite{Bloom-2011,Burrows-2011,Zauderer-2011,Levan-2011}. The ``fundamental
plane'' of the black hole accretion, described by a relation between radio luminosity, X-ray
luminosity, and black hole mass may be used to determine mass from the observed radio and
X-ray luminosities. This way, \cite{Miller-2011} estimate the black hole mass of Sw
164449+57: to be $\log(M/M_\odot) = 5.5 \pm 1.1$.

{\cite{Reis-2012} noticed that Swift J1644 is a beamed source and therefore its accretion
disk must be viewed at very low inclination (nearly face-on). This excludes the possibility
that the disk oscillations are modulated directly by the relativistic Doppler effect and by
the light trajectory bending. However, these effects work only for highly inclined disks
(\cite{Bursa-2004}). It is not obvious what the modulation mechanism  is  in the case of the
black hole disks that are seen nearly face-on, in particular whether the disk oscillations
may modulate (with the same frequencies) the properties of the disk. This is an interesting
issue for further theoretical studies. In this context we would like to note that
\cite{McKinney-2012} discuss the possibility of exciting QPOs through a disk-jet coupling
(see also \cite{Krolik-2012} who discuss the jet issue specifically for Swift J1644.)}

We conclude that the hypothesis that the QPO frequency observed in Sw 164449+57 is one of
the twin peak 3:2 frequencies, and the hypothesis of the high spin in this source, are
together consistent with the above-mentioned low mass estimates ($M \sim 10^5\,M_\odot$),
implying an intermediate mass black hole in Sw 164449+57.

\begin{acknowledgements}
{We are grateful to the referee who quickly wrote a very helpful report and who pointed out
a few issues connected with the low inclination of the disk in Swift J1644.} This work was
done during a visit of M.A.A. to Kavli Institute of Astronomy and Astrophysics and Astronomy
Department at Peking University, and was supported by the National Natural Science
Foundation of China (NSFC11073002) and China Scholarship Council (2009601137) as well as by
M.A.A. grants: IAU travel grant, Polish NCN UMO-2011/01/B/ST9/05439 grant, and Swedish VR
grant.
\end{acknowledgements}


\end{document}